\documentclass[prd,aps,showpacs,amsmath,amssymb,twocolumn,nofootinbib]{revtex4-1}
\usepackage{graphicx}
\usepackage{dcolumn}
\usepackage{bm}

\begin{document}
\title{Stability of thin-shell interfaces inside compact stars}

\author{Jonas~P.~Pereira$^{1,2}$, Jaziel~G.~Coelho$^{1,3}$, Jorge A. Rueda$^{1,3,4}$}
\email{jonaspedro.pereira@icranet.org\\jaziel.coelho@icranet.org\\jorge.rueda@icra.it}
\affiliation{$^1$Dipartimento di Fisica and ICRA, Sapienza Universit\`a di Roma, P.le Aldo Moro 5, I--00185 Rome, Italy}
\affiliation{$^2$Universit\'e de Nice Sophia Antipolis, 28 Av. de Valrose, 06103 Nice Cedex 2, France}
\affiliation{$^3$ICRANet, Piazza della Repubblica 10, I--65122 Pescara, Italy}
\affiliation{$^4$ICRANet-Rio, Centro Brasileiro de Pesquisas F\'isicas, Rua Dr. Xavier Sigaud 150, Rio de Janeiro, RJ, 22290--180, Brazil }
\date{\today}

\begin{abstract}
We use the thin-shell Darmois-Israel formalism to model and assess the stability of the interfaces separating phases, e.g. the core and the crust, within compact stars. We exemplify the relevance and non-triviality of this treatment in the simplest case of an incompressible star, in constant pressure phase transitions, and in the case of strange quark stars with crust.
\end{abstract}

\maketitle

\section{Introduction}

In the Newtonian theory of gravity, it is well-known the procedure to deal with a given surface of discontinuity. One should simply impose the continuity of the gravitational potential across it and the discontinuity of the gravitational field comes from its surface mass. Such boundary conditions can be easily concluded from the (linear) field equation, given the precise notion of reference systems. Nevertheless, in general relativity the problem is much more involved, due to the nonlinearity of the field equations and the principle of general covariance \cite{1966NCimB..44....1I,1967NCimB..48..463I}. For the elucidation of the problem and references, see Refs. \cite{2005CQGra..22.4869L,1966NCimB..44....1I,1967NCimB..48..463I}. The solution to this problem consists of imposing
specific boundary conditions to the induced metric tensor and the extrinsic curvature \cite{1973grav.book.....M,2004reto.book.....P} on a hypersurface splitting spacetimes in a manifestly covariant way. Such a procedure is generically called either the thin-shell formalism or the Darmois-Israel formalism. It has been applied to a variety of scenarios, assessing the physical properties of e.g.~dynamic thin-layers \cite{1990CQGra...7..585F,1991PhRvD..44.1891B,1995PhRvD..52.7318P,2002PhRvD..66h4021G,2003gr.qc.....8025K,2005CQGra..22.4869L,2007PhRvD..75b7501B,2011PhRvD..83j4009E,2012PhRvD..86h4004H, 2013PhLA..377.2450R,2013PhRvD..87f4041E}, quantum fields in thin-shell spacetimes \cite{2013PhRvD..87j4039L,2014PhRvD..90d4053M}, wormholes \cite{2005CQGra..22.4869L,2010PhRvD..82h4023D}, and radiating spheres \cite{1989PhLA..138....1H, 1989ApJ...339..339H,1991PhRvD..44.2286D, 1992PhRvD..46.2723A,1992PhRvD..45.3341E,1993PhRvD..48.2961E}. It was already shown that such a formalism is equivalent to searching for distributional solutions to Einstein's equations \cite{1996JMP....37.5672M}. For details about the derivation of the hypersurface conditions in this scenario, see Ref.~\cite{2004reto.book.....P}.

The thin-shell formalism would be meaningful for stars that are expected to display interfacial layers much smaller than their characteristic sizes \cite{1996csnp.book.....G}, endowed generically with nontrivial quantities, such as surface tensions and surface energy densities. Some systems of our interest in this line can be, for example, compact stars with interfaces separating their cores and their crusts, e.g.~strange quark stars and neutron stars. We highlight that the Darmois-Israel formalism gives the nontrivial properties of transitional layers fully taking into account general relativity, but just under the macroscopic point of view. More generally, the thin-shell formalism would be the proper formalism for approaching any gravitational system that presents discontinuous behaviors in their physical parameters.

We consider here a star with different phases as the match of given spacetimes split by hypersurfaces that could possess nontrivial properties, which in turn lead to nontrivial dynamics, imputable to general relativity. The dynamics of a shell does not imply compression or dilution of the matter contents in the spacetimes, since it is induced by fixed geometries associated with its adjacent spacetimes. When a shell moves, one phase (thick layer) tends to ``swallow'' the other, leading the shell to ``absorb'' degrees of freedom to it \cite{2007arXiv0710.3193C}, in order to compensate the changes of the energy-momentum in the adjacent phases and on itself. All of these aspects shall be clarified subsequently.

Duly taking into account surface degrees of freedom in a stratified astrophysical system is of cardinal importance because they guarantee that the global spacetime represented by the union of the spacetime solutions matched at given interfaces is also a solution of the general relativistic equations. The consistency of the match of spacetimes is far from trivial owing to the intrinsic non-linearity of the Einstein's general theory of relativity.

Our aim in this article is to analyze, within the thin-shell formalism, the stability of transitional/interfacial thin layers present in compact stars
against radial displacements of the shell. These latter perturbations are the outcome of surface ones only. They are assumed to occur adiabatically and therefore will propagate with the speed of the sound \cite{1959flme.book.....L,1950RSPSA.200..248C}, to be properly inspected for these continuous systems. In order to evidence the intrinsic role that the surface degrees of freedom (surfaces tensions and surface energy-densities) play in stars with discontinuities, we
shall work in the context where just surface perturbations are present. This analysis is believed to be important since instabilities of interfaces in compact stars (triggered by physical mechanisms we shall not investigate) might be a direct sign of instabilities of the whole systems, even in the absence of perturbations in the glued spacetimes. Besides, it would pave the way for stability analyses where the phases of the system are also perturbed.

The article is organized as follows. In section \ref{overview} we introduce the general equations of the thin-shell formalism, while in section~\ref{interpretation} we make a physical interpretation of the shell's parameters leading to a general relativistic extension of the concepts of surface energy-density, surface tension, and the Young-Laplace equation of the mechanical equilibrium of phase-separating interfaces. The equation of motion of the thin-shell and the condition for the shell's stability are derived in section~\ref{sec:stability}. The consequences of the treatment for the case of constant-pressure phase-transitions are outlined in section~\ref{constpress}. We present in section~\ref{numerical} the application of the formalism to the simplest case of stars made of incompressible matter and to the case of strange quark stars with crust. Section~\ref{slowrot} shows the extension of the thin-shell treatment of an interface in the case of slow rotation. Finally, in section~\ref{disc} we outline and discuss the main conclusions of the article. We use geometric units $G=c=1$ throughout the article. Unless it is not otherwise stated, we work with a signature $+2$. Greek indexes run from zero to three, while Latin ones run from zero to two.

\section{Thin-shell formalism in the spherically symmetric case}\label{overview}

The Darmois-Israel formalism can be enunciated as follows \cite{2005CQGra..22.4869L}.
Consider two pseudo-Riemannian manifolds, $M_{+}$ and $M_{-}$, endowed with metric fields $g_{\alpha\beta}^{+}(x^{\mu}_{+})$ and
$g_{\alpha\beta}^{-}(x^{\mu}_{-})$, with respect to two independent coordinate systems $x^{\mu}_{+}$ and $x^{\mu}_{-}$.
Assume that such manifolds have boundaries $\Sigma_{+}$ and $\Sigma_{-}$. If such boundaries are identified, then a natural match of
manifolds can be done, where the resultant manifold, $M$, is the union of the aforementioned ones. Call such a common hypersurface as $\Sigma$.
Assume that a coordinate system  $y^{a}$ is adapted to it. As any hypersurface, it represents a constraint of the
spacetime coordinates, here defined as $\Psi^{\pm}(x^{\mu}_{\pm})=0$. It can also be written in the parametric form $x^{\mu}_{\pm}=x^{\mu}_{\pm}(y^{a})$.
A natural basis can be defined on $\Sigma$ by means of tangent vectors to its coordinate curves. Define
the components of it as $e^{\mu\pm}_{a}\doteq \partial x^{\mu}_{\pm}/\partial y^{a}$. With these basis vectors, one can easily find the
induced metric on $\Sigma$, $h_{ab}$, when the spacetime line element is constraint to such a hypersurface.
From our previous reasoning, it is clear that such an induced geometry must be unique. Indeed, this is the first boundary conditions one has to
impose in order to have a well-defined pseudo-Riemannian manifold made out of the glue of two other ones. This can be viewed as the general relativistic generalization of the continuity of the gravitational potential across a surface in the Newtonian theory of gravity.
The geometry of $\Sigma$ is
\begin{equation}
h_{ab}=g_{\mu\nu}^{\pm}e^{\mu\pm}_{a}e^{\nu\pm}_{b}\label{indmetr},
\end{equation}
which is independent of the coordinate systems $x^{\mu}_{\pm}$ utilized.

The normal unit four-vector to $\Sigma$ is defined such that
\begin{equation}
n_{\mu}^{\pm}\doteq \frac{\epsilon\partial_{\mu}\Psi^{\pm}}{|g_{\pm}^{\alpha\beta}\partial_{\alpha}\Psi^{\pm}\partial_{\beta}\Psi^{\pm}|^{\frac{1}{2}}}\label{normal},
\end{equation}
where $\partial_{\mu}\doteq {\partial}/{\partial x^{\mu}}$ and it is tacit that $x^{\mu}$ is actually a shortcut to $x^{\mu}_{\pm}$. Besides,  $n^{\alpha}n_{\alpha}=\epsilon=\pm 1$, depending on the nature of the hypersurface. Notice that the case where $n_{\alpha}$ is null is not contemplated here. Equation (\ref{normal}) also guarantees that $n^{\mu}\partial_{\mu}\Psi>0$.

Another important quantity for characterizing a hypersurface is its extrinsic curvature (or second fundamental form)
\begin{equation}
K_{ab}\doteq n_{\alpha ; \beta}e^{\alpha}_{a}e^{\beta}_{b}\label{extcurv},
\end{equation}
where we did not put the ``$\pm$'' labels just not to overload the notation. One sees that the extrinsic curvature components are the tetrad decomposition of the tensor $n_{\mu;\nu}$, thence a tangent vector \cite{2004reto.book.....P}.

Let us define the jump of a given tensorial quantity across $\Sigma$ as $[A]^+_-\doteq A(x^{+})|_{\Sigma}-A(x^{-})|_{\Sigma}$. It is implicit in the previous
definition that $A(x^{\pm})|_{\Sigma}$ stands for a given quantity $A$ being evaluated in arbitrary points belonging to the disjoint regions implied by $\Sigma$ and then taken the limit when they tend to an arbitrary point on $\Sigma$. The energy-momentum tensor on $\Sigma$ coming from general relativity that guarantees a distributional solution to the field equation, $S^{\mu\nu}$, can be expressed in terms of the jump of the extrinsic curvature by means of the Lanczos equation \cite{1966NCimB..44....1I,1996JMP....37.5672M}
\begin{equation}
S_{\mu\nu}=S_{ij}e_{\mu}^{i}e_{\nu}^{j},\;S_{ij} = -\frac{\epsilon}{8\pi}([K_{ij}]^+_- - h_{ij}[K]^+_-)\label{lanczos},
\end{equation}
where $K\doteq h^{lm}K_{lm}$. This is the second boundary conditions one should impose. Clearly, it is the (manifestly covariant) generalization of the jump the gravitational field experiences in the Newtonian theory. Note that $S^{\mu\nu}$ also contributes to the energy-momentum tensor of the matched manifold as $S^{\mu\nu}\delta(\Psi)$, $\delta$ the Dirac delta, and this is essential to guarantee the validity of the Bianchi identities for $M$ \cite{1996JMP....37.5672M}.
The evolution equation to $S_{ij}$ is \cite{2004reto.book.....P}
\begin{equation}
S^{a}{}_{b|a}=-\epsilon[T_{\alpha\beta}e^{\alpha}_{b}n^{\beta}]^+_-\label{evoequ},
\end{equation}
where $A_{ab|c}\doteq A_{\mu\nu;\alpha}e^{\mu}_{a}e^{\nu}_{b}e^{\alpha}_{c}$ \cite{2004reto.book.....P}. One sees from the above equation that thin-shells in continuous systems naturally are subjected to current fluxes, generically given by $j_{a}=T_{\alpha\beta}e^{\alpha}_{a}n^{\beta}$ \cite{2004reto.book.....P}. Notice that thin-shells gluing vacuum spacetimes do not present such currents.

For spherically symmetric spacetimes, following Ref.~\cite{2005CQGra..22.4869L}, we take the line element of the glued spacetimes as
\begin{equation}
ds^2_{\pm}=-e^{2\alpha_{\pm}(r_{\pm})}dt^2_{\pm}+e^{2\beta_{\pm}(r_{\pm})}dr^2_{\pm}+r^2_{\pm}d\Omega^2_{\pm}\label{linelsph},
\end{equation}
where
\begin{equation}
d\Omega^2_{\pm}= d\theta^2_{\pm}+
\sin^2\theta_{\pm}d\varphi^2_{\pm}\label{solangle}.
\end{equation}
We consider that the hypersurface $\Sigma$ is described by the equation $\Psi_{\pm}= r_{\pm}-R(\tau)$=0, where $\tau$ is the proper time of an observer on it. Besides, we take $\theta_{\pm}=\theta$ and $\varphi_{\pm}=\varphi$ for the remaining coordinates on $\Sigma$. In other words, we are selecting a geodetic observer for describing the geometry of the thin-shell. For the above choice of coordinates, it is clear that $e^{\mu\pm}_{0} = U^{\mu}_{\pm}$, the four-velocity of $\Sigma$, hence, $n_{\mu}U^{\mu}=0$. The geometry of $\Sigma$ is only well defined (or unique) when
\begin{equation}
\dot{t}_{\pm}=e^{\beta_{\pm}-\alpha_{\pm}}\sqrt{\dot{R}^2+e^{-2\beta_{\pm}}}\label{tdot},
\end{equation}
where we are defining the dot operation as the derivative with respect to $\tau$. Taking into account the previous points, the
geometry of $\Sigma$ is therefore
\begin{equation}
ds^2_{\Sigma}=-d\tau^2+R^2(\tau)(d\theta^2+\sin^2\theta d\varphi^2)\label{Sgeom}.
\end{equation}

The Lanczos equation (\ref{lanczos}) for the spherically symmetric case implies that \cite{2005CQGra..22.4869L}
\begin{equation}\label{Sab}
S^{a}{}_{b}= {\rm diag}(-\sigma,{\cal P}, {\cal P}),
\end{equation}
with
\begin{equation}
\sigma = -\frac{1}{4\pi R}\left[\sqrt{e^{-2\beta}+\dot R^2}\right]^+_-\label{enerdens},
\end{equation}
\begin{equation}
{\cal P}= -\frac{\sigma}{2} + \frac{1}{8\pi R}\left[\frac{R\alpha'(e^{-2\beta}+\dot R^2)+\ddot R R + \beta' R\dot R^2}{\sqrt{e^{-2\beta}+\dot R^2}}\right]^+_-\label{surfdens},
\end{equation}
where the prime was defined as the derivative with respect to the radial coordinate defined in each region of the glued manifold $M$.
Eqs.~(\ref{evoequ}) and (\ref{Sab}) for this case give \cite{2005CQGra..22.4869L}
\begin{equation}
\dot\sigma=-\frac{2 \dot R}{R}(\sigma + {\cal P})+\Delta \dot R\label{evolequ},
\end{equation}
with
\begin{equation}
\Delta\doteq \frac{1}{4\pi R}\left[(\alpha'+\beta')\sqrt{e^{-2\beta}+\dot R^2}\right]^+_-\label{Delta}.
\end{equation}
Eq.~(\ref{evolequ}) can be rewritten in the much more appealing form
\begin{equation}
\frac{d}{d\tau}(4\pi R^2 \sigma)=-\left({\cal P}-\frac{\Delta R}{2} \right)\frac{d}{d\tau}(4\pi R^2)\label{thermequ},
\end{equation}
resembling a first law of thermodynamics for the spherically symmetric surface. This then would lead us to the interpretation of $\sigma$ as the energy-density on $\Sigma$, while ${\cal P}$ as the pressure (surface tension) connected with the work done by the internal forces in the shell. In the next section we shall see that this is indeed the case. Besides, ${4\pi R^2\dot{R}\Delta}$ is the work done by the nonzero normal flux of momentum $T_{\alpha\beta}U^{\beta}n^{\alpha}$ across $\Sigma$.

We would like also to emphasize that the thin-shell formalism in the spherically symmetric case leads to Eqs.~(\ref{enerdens}) and (\ref{surfdens}) [or
Eq.~(\ref{thermequ})], while the unknown variables to the problem are $\sigma$, ${\cal P}$ and $R$. This means that an equation of state ${\cal P}={\cal P}(\sigma)$, must be given for closing the system of equations. Such an equation of state would embrace the microphysics of the matter inside the shell.
Otherwise, a free parameter will be present into the formalism.

It is worth mentioning that the above equations are general in the spherically symmetric case and can be applied to configurations with any
energy-momentum tensor consistent with such an assumption. This therefore includes configurations endowed with a non-vanishing electric (but not magnetic) field, $E$. The total energy-momentum tensor is in such a case the sum of the isotropic matter energy-momentum tensor, $(T^\alpha_\beta)_{\rm matter}$ $=$ ${\rm diag}(-\rho,P,P,P)$ and the anisotropic electrostatic energy-momentum Maxwell tensor, $(T^\alpha_\beta)_{\rm elec}$ $=$ $(8\pi)^{-1}E^2{\rm diag}(-1-,1,1,	1)$, leading to a total radial pressure, $P_r=P-(8\pi)^{-1}E^2$, different from the resultant tangential pressures, $P_\bot\doteq P_{\theta\theta}$ $=$ $P_{\varphi\varphi}=P+(8\pi)^{-1}E^2$. However, the energy-momentum tensor of the hypersurface holds still the perfect-fluid form (\ref{Sab}) since the pressure anisotropy exists only in the radial direction, dimension which is suppressed by definition in the thin-shell treatment of an interface.

\section{Physical interpretation of the thin-shell parameters}
\label{interpretation}

In this section we interpret the quantities $\sigma$ and ${\cal P}$ arising from the thin-shell formalism for the spherically symmetric case. For the sake of simplicity we do not consider the presence of electric charge. The Einstein's equations describing in this case can be written as \cite{1975ctf..book.....L}
\begin{equation}\label{g11equ}
e^{-2\beta(r)}\doteq 1-\frac{2m(r)}{r},\;\; m(r)=4\pi\int_{0}^{r}\rho(\bar{r})\bar{r}^2d\bar{r},
\end{equation}
\begin{equation}\label{derg00}
\alpha'=\frac{8\pi p r^3+2m(r)}{2 r\left\{r-2m(r)\right\}},
\end{equation}
where we defined $\rho$ and $p$ as the energy-density and the radial pressure, respectively, as measured by local Lorentz observers. Here we suppressed the $\pm$ labels anew just not to overload the notation. It is clear that $\sigma$ and ${\cal P}$ must be relativistic generalizations to classical quantities. In this regard, we proceed with their weak field analysis for an equilibrium configuration at $r=R_0$. This is done by assuming that $m_{\pm}(R_0)/R_0\ll 1$. For this case, it is easy to show that Eqs.~(\ref{enerdens}) and (\ref{surfdens}), on account of Eqs.~(\ref{g11equ}) and (\ref{derg00}), read
\begin{equation}
\sigma_0=\frac{1}{4\pi}[g(R_0)]^+_-,\;\; g(R_0)\doteq \frac{m(R_0)}{R_0^2}\label{Newtenergdens},
\end{equation}
\begin{equation}
{\cal P}_0=\frac{1}{2}R_0 [p(R_0)]^+_-\label{YoungLap}.
\end{equation}
We see from Eq.~(\ref{Newtenergdens}) that $\sigma$ is indeed the the generalization of the surface energy-density (mass density) of the shell. Eq.~(\ref{YoungLap}) is the well-known Young-Laplace equation for the mechanical equilibrium of a spherically symmetric bubble-like system \cite{2003EJPh...24..159R} and hence $\cal P$ can be identified as the general relativistic generalization of the surface tension. For the latter case, we notice that in the lowest order of approximation, the surface tension is just obtained by geometric considerations as in the Young-Laplace approach. For higher order corrections, once more from Eqs.~(\ref{g11equ}) and (\ref{surfdens}) in the static case, we have
\begin{equation}
{\cal P}_0= \frac{1}{2}R_0 [p]^+_- + \frac{G}{16\pi R_0^3}[m^2]^+_- + \frac{G[p\,m]^+_-}{2c^2} + \frac{G^2[m^3]^+_-}{8 \pi c^2 R_0^4} + ... \label{gravindusurfdens},
\end{equation}
where we have restored the units for completeness. From the above expression, it is manifest the appearance of gravitational and general relativistic corrections to the surface tension. We stress that the above well-known classical results for $\sigma$ and ${\cal P}$ are a direct consequence of the Israel-Darmois formalism. Hence, any general relativistic system endowed with nontrivial surface quantities must be described by the aforementioned method.

\section{Stability of the thin-shell against radial perturbations}\label{sec:stability}

We now proceed with the stability analyses of the thin-shell against radial perturbations
(the description where also the adjacent spacetimes are perturbed will be analyzed elsewhere). We start by rewriting Eq.~(\ref{enerdens})
in the suggestive form \cite{2005CQGra..22.4869L}
\begin{equation}
V(R)+\dot R^2=0\label{shelldyn},
\end{equation}
where we have introduced the shell's effective potential
\begin{equation}
V(R)\doteq \frac{1}{2}(e^{-2\beta_-}+e^{-2\beta_+})-\frac{1}{4}(4\pi R \sigma)^2-\frac{1}{4}\left(\frac{[e^{-2\beta}]^+_-}{4\pi R \sigma} \right)^2\label{effpot}.
\end{equation}
The solution, $R(\tau)=R_0=$constant, implies that $V(R_0)=V'(R_0)=0$, that in turn leads $R_0$ to be automatically a critical point to the effective potential. Assume now small radial displacements from this solution. As it is well-known, just in the case $V''(R_0)>0$ one has a stable behavior of the system. From Eqs.~(\ref{effpot}), (\ref{evolequ}) and the identity $A'|_{\Sigma}=(\dot A/ \dot R)|_{\Sigma}$, after some simple calculations, one shows that the stability condition, for the case $\sigma>0$, can be written as
\begin{equation}
{\tilde V}'' > 0,\label{stabcond}
\end{equation}
where
\begin{equation}
{\tilde V}'' \doteq -[e^{-\beta} \left\{(2\eta +1)(1+R_0\beta')-R_0^2(\alpha''-\beta'\alpha')\right\}]^+_-,
\end{equation}
being
\begin{equation}
v^2_s\doteq \eta \doteq \frac{\partial {\cal P}}{\partial \sigma},
\end{equation}
as we show in the appendix~\ref{app:sound}, the squared of the speed of the sound in the shell.

We are not interested here in exploring the microphysics of the shell. Thence, we will allow $\eta$ to be a free parameter in our description. Nevertheless, it must be borne in mind that a physical system is ascribed solely to an equation of state (therefore a single value of $\eta$ for a given point) and our analyses with free $\eta$ can be seen as the construction of a generic stability catalog for given matched spacetimes. It is worth mentioning that the issue of equations of state for surfaces remains thus far knotty even in the forefront investigations of material sciences, where there are yet phenomenological models awaiting theoretical frameworks \citep{surfaceforces2011}.

By requesting the stability condition (\ref{stabcond}), we shall constrain $\eta$, as well as other parameters appearing in $\beta_{\pm}$ and $\alpha_{\pm}$ that particularize the configuration, so that they lead to stable solutions to the thin-shell. In doing so, we shall impose that the speed of the sound in the shell does not exceed the speed of light, i.e. $|v_s|\leq c$.

\subsection{Newtonian limit of the stability condition}

It is instructive to analyze the above stability condition [Eq.~(\ref{stabcond})] in the Newtonian limit. The relevant equations to be taken
into account here are Eqs.~(\ref{g11equ}) and (\ref{derg00}) in the weak field limit, where one can make the approximations
\begin{equation}
\beta\simeq \frac{m(r)}{r}\;\; \mbox{and}\;\; \alpha' \simeq \phi' = \frac{m(r)}{r^2} \label{g11g00clas},
\end{equation}
where $\phi$ is the gravitational potential, such that the gravitational field is $\vec{g}=-\nabla \phi$.
We are also assuming that $m(r)/r\ll1$. When one substitutes Eq.~(\ref{g11g00clas}) into Eq.~(\ref{stabcond}),
one concludes that the stability at $R_0$ is translated into
\begin{equation}
2\eta \left[m'(R_0)-2\frac{m(R_0)}{R_0} \right]^+_- = 2\eta R_0^2 [\phi'']^+_- <0 \label{stabpert}.
\end{equation}
That the stability condition is related to the second derivative of a potential is self-explanatory. Its jump being negative at a surface of discontinuity means that the norm of the gravitational force density must decrease, as one excepts from stable systems. The Poisson's equation for the potential given by Eq.~(\ref{g11g00clas}) reads
\begin{equation}
\frac{m'(r)}{r^2}= 4\pi \rho (r)\label{poisson},
\end{equation}
with $\rho(r)$ the mass density at $r$. Putting Eq.~(\ref{poisson}) into Eq.~(\ref{stabpert}), the latter can be cast as
\begin{equation}
4\eta \left[2 \pi \rho (R_0) R_0^2-\frac{m(R_0)}{R_0} \right]^+_-<0\label{irrstabpert}.
\end{equation}
From the above equation and previous considerations on $\eta$, one sees that thin-shells with no surface mass densities are stable if $[\rho(R_0)]^+_-<0$. This is exactly what one intuitively expects for stars. For vacuum systems, the stability simplifies to $[m(R_0)]^+_->0$, i.e., the surface mass on $R_0$ should be positive. The physical reason for having stability even for a shell embedded in vacuum spacetimes is due to the induced gravitational surface tension, as it can be seen from Eq.~(\ref{gravindusurfdens}).

\section{Interfaces at constant pressure}\label{constpress}

We turn now to show a first immediate consequence of the thin-shell formalism: an astrophysical body with an interface splitting two phases under
a constant pressure is stable against radial displacements whenever the mass the interfacial shell nests is much smaller than the total mass of
the system. The hypotheses imply that $[p]^+_-=0$ and $\sigma=0$, which leads to $[e^{\beta}]^+_-=0$.
For this case, Einstein's equations on the hypersurface $\Sigma$ give
\begin{equation}
[\beta']^+_-=4\pi e^{2\beta}R [\rho]^+_-\;\; \mbox{and}\;\; [\alpha']^+_-=0 \label{disccondp0}.
\end{equation}
In the static case, from Eqs.~(\ref{surfdens}) and (\ref{disccondp0}), we have that ${\cal P}=0$. The aforesaid hypotheses do not render the system continuous since $[\rho]^+_-$ is yet unspecified. In the dynamic case generally ${\cal P} \neq 0$. From Eq.~(\ref{stabcond}) and the above equation, the stability condition becomes
\begin{equation}
2\eta R_0 [\beta']^+_-<0\label{stabcondp0}.
\end{equation}
We have shown that $\eta$ is the square speed of the sound and therefore it must be positive. Hence, from Eqs.~(\ref{disccondp0}) and (\ref{stabcondp0}) we see that the hypersurface is stable iff $[\rho]^+_-<0$, as one expects from a physically reasonable configuration with a monotonically decreasing energy-density with the distance. Therefore, we have generically shown that the interface of a system separating a constant pressure phase-transition with negligible interfacial mass is always stable against radial perturbations. This result is therefore applicable to the stars with constant pressure phase-transitions, either in the scope of the Maxwell or Gibbs phase-transition constructions. It is important to recall, however, that in systems with more than one conserved charges (e.g. baryon and electric) the Gibbs construction leads to the appearance of mixed phases, in between of the pure phases, with an equilibrium pressure that varies with the density. This may lead, in turn, to a spatially extended phase-transition of non-negligible thickness with respect to the star radius (see, e.g., \cite{1996csnp.book.....G}, and references therein). It is clear that such an extended mixed phase region, separating the two pure phases, cannot be treated within a thin-shell approach. Nevertheless, its interfaces demarcating the onset and the termination of a mixed phase always can, as well as thin mixed phases. For the aforementioned constructions, in a sense the thin-shell treatment is  more suitable to model configurations in which the phase-transition follows a Maxwell construction, where the phases are in ``contact'' each other\footnote{There is yet a debate in the literature concerning the use of Maxwell or Gibbs constructions for thermodynamic phase-transitions in multicomponent systems such as the ones present in compact stars because they lead basically to the same results for the masses and radii of neutron stars \cite{2010EPJA...46..413P}.}. It is worth mentioning that these treatments of the existing phases in compact stars subject the system to the condition of local charge neutrality, and so they do not account for the possible presence of interior Coulomb fields. Indeed, the complete equilibrium of the multicomponent fluid in the cores of compact stars needs the presence of an electric charge separation caused by gravito-polarization effects \cite{Rotondo:2011rj,Rueda:2011nq}, favoring a sharp core-crust transition that ensures the global, but not the local, charge neutrality \cite{2012NuPhA.883....1B,2014NuPhA.921...33B}. In order to keep the presentation of the applications apropos of the thin-shell scenario as simple as possible, we consider hereafter the configurations in the limit where the system is locally neutrality, leaving the more complex case of global charge neutrality to be treated elsewhere.

\section{Some specific examples of thin-shell interfaces in compact stars}\label{numerical}

\subsection{Incompressible stars with interfaces}

In order to gain more intuition about the main aspects of the thin-shell formalism, it is instructive to work first with an exact fully relativistic case. In this
regard we analyze in this section the stability of stars with constant densities that present ``phase transitions''. Let us assume a star that has a constant density $\rho^{-}$ from the origin to a radius $R$ (that could be even dynamic), and from $R$ until its surface $R_s$ it has another constant density $\rho^{+}$. Assume further that its associated discontinuity surface $\Sigma$ has a negligible energy-density when compared to either regions it defines. We shall seek solutions that are regular at $r=0$. The integration of the Einstein's equations with the aforesaid assumptions lead us to the following solutions. For $r<R$:
\begin{equation}
e^{-2\beta^{-}}=1-\frac{8\pi r^2\rho^{-}}{3},\;\; e^{2\alpha^{-}}=A_0^-\left( \frac{\rho^- + p^- }{\rho^- + p_0^-}\right)^{-2}\label{cdinnermetric},
\end{equation}
where $A_0^-$ and $p_0^-$ are arbitrary constants of integration and the inner pressure $p^-$ is
\begin{equation}
p^-= \rho^- \frac{\left\{(\rho_- + 3p_0^-) \left( 1-\frac{8 \pi \rho^- r^2 }{3}\right)^{\frac{1}{2}}-(\rho^- + p_0^-) \right\}}{3({\rho^- + p_0^-})- (\rho^- + 3p_0^-) \left( 1-\frac{8 \pi \rho^- r^2 }{3}\right)^{\frac{1}{2}}}\label{cdinnerpres}.
\end{equation}
For $R<r\leq R_s$,
\begin{eqnarray}
e^{-2\beta^{+}}&=& 1-\frac{8 \pi}{3 r}\left\{ R^3(\rho^- - \rho^+) + \rho^+ r^3 \right\}\label{cdouterg11},\\
e^{2\alpha^{+}}&=& A_0^+\left( \frac{\rho^+ + p^+ }{\rho^+ + p_0^+}\right)^{-2}\label{cdouterg00},
\end{eqnarray}
with
\begin{widetext}
\begin{equation}
A_0^+=\left(1-\frac{2 M}{R_s} \right) \left(\frac{\rho^+}{\rho^+ + p_0^+} \right)^2\label{cdouterA0g00},
\end{equation}
\begin{equation}
p^+ = \rho^+ \frac{(\rho^+ + 3p_0^+) \left( 3 - 8 \pi \rho^+ r^2\right)^{\frac{1}{2}}-(\rho^+ + p_0^+)(3 - 8 \pi \rho^+ R^2)^{\frac{1}{2}}}{3(\rho^+ + p_0^+)(3 - 8 \pi \rho^+ R^2)^{\frac{1}{2}} - (\rho^+ + 3p_0^+)\left(3 - 8 \pi \rho^+ r^2 \right)^{\frac{1}{2}}}\label{cdouterpres},
\end{equation}
\begin{equation}
p_0^+ = \rho^+ \frac{\left(3 - 8 \pi \rho^+ R_s^2\right)^{\frac{1}{2}}-(3 - 8 \pi \rho^+ R^2)^{\frac{1}{2}}}{(3 - 8 \pi \rho^+ R^2)^{\frac{1}{2}}- 3\left(3 - 8 \pi \rho^+ R_s^2\right)^{\frac{1}{2}} }\label{cdouterpres0}.
\end{equation}
\end{widetext}
The total mass of the system was defined such that
\begin{equation}
M\doteq \frac{4\pi}{3} \left\{ R^3(\rho^- - \rho^+) + \rho^+ R_s^3\right\}\label{cdtmass}.
\end{equation}
In the above equations, it was assumed that the outer pressure at $R_s$ vanishes. Eq.~(\ref{cdouterA0g00}) guarantees the outside match of the star with the Schwarzschild metric. In the scope of the stability of a thin-shell immersed in a continuous system, the constant multiplicative factor on the time-time metric component is not of importance. This is related to the freedom in re-scaling the time coordinate for the metric in Eq.~(\ref{linelsph}). Therefore, $A_0^{\pm}$ will not play any relevance to our stability analyses. Notice further that we can also have solutions with $p^+(R)=p^-(R)$, by properly adjusting the arbitrary constant of integration $p_0^-$. As we already know from the preceding section, this case is stable iff $[\rho]^+_-<0$. Let us analyze another case, where $p_0^-$ is a free parameter.

It is convenient to relate $R$ and $R_s$, as well as $\rho^{\pm}$. Let us assume that $R_s=C_1 R$ and $\rho^{-}= C_2\rho^+$. We stress the fact that the solution for constant densities lead to the constraint of $\rho R_s^2$ to be of the order of unity. Assuming that the radii of our systems are similar to the ones expected to neutron stars, reasonable values for the $r$-coordinate would be of the order of $10^{6}$~cm. Therefore, the maximum densities allowed for constant densities stars would be $\rho_{\rm max}\simeq 10^{-12}$~cm$^{-2}= 10^{16}$~g~cm$^{-3}$. Such densities are well above the nuclear one, of the order of $10^{14}$~g cm$^{-3}\simeq 10^{-14}$~cm$^{-2}$. The pressure at the origin $p_0^-$ for this stratified system is arbitrary. Nevertheless, reasonable values for it are of the order of (or higher than) $p_0^- \simeq 10^{35}$~dyn cm$^{-2}\simeq 10^{-13}$~cm$^{-2}$.

\begin{figure}[!htbp]
	\centering
		\includegraphics[width=\hsize,clip]{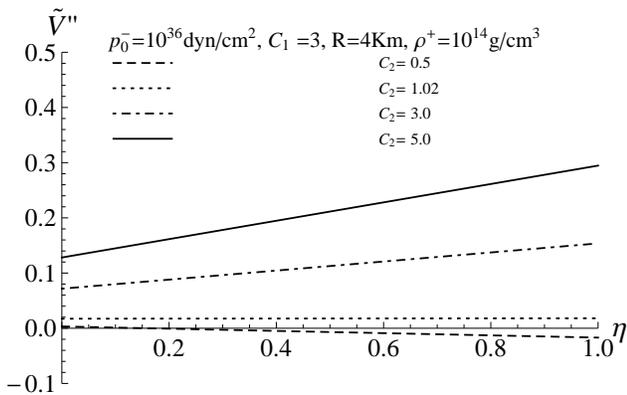}
		\caption{Stability of thin-shells as a function of the square of the speed of sound $\eta$, for various values of the parameter $C_2$. One sees here that for the choice of the remaining parameters close to the ones of ordinary neutron stars, not any configuration for the system would lead to stable solutions. Interestingly, the case $[\rho]^+_- > 0$ (dashed curve) gives a window of stable solutions, contrary to what is expected from the classical Newtonian case, where stable solutions are only possible with negative jump of the mass density. }\label{cd106}
\end{figure}

\begin{figure}[!htbp]
	\centering
		\includegraphics[width=\hsize,clip]{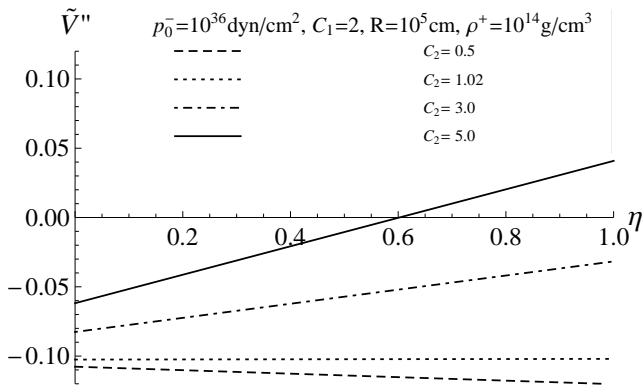}
		\caption{Stability of thin-shells as a function of the square of the speed of sound $\eta$, for various values of the parameter $C_2$. One sees here that the change of the choice of the parameters $C_2$ and $R$ as related to Fig.~\ref{cd106} modifies the stability of the system.}\label{cd105}
\end{figure}

\begin{figure}[!htbp]
	\centering
		\includegraphics[width=\hsize,clip]{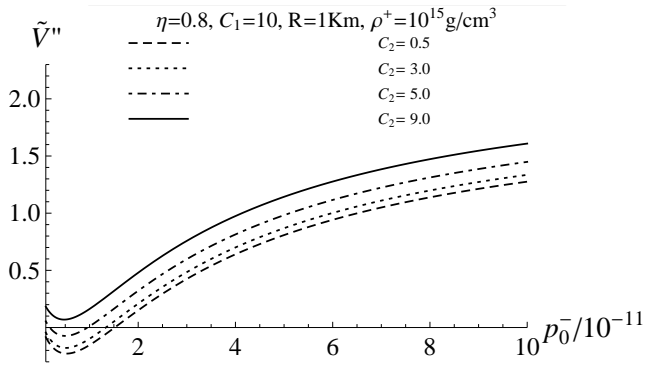}
		\caption{Stability of thin-shells as a function of the central pressure $p_{0}^{-}$ (in units of $cm^{-2}$), for various values of the parameter $C_2$. For the parameters chosen, the change of $C_2$ and $R$ influence the stability for a given equation of state, as exemplified here by $\eta=0.8$.}\label{cdp0}
\end{figure}

By replacing Eqs.~(\ref{cdinnermetric}), (\ref{cdouterg11}) and (\ref{cdouterg00}) in Eq.~(\ref{stabcond}), we would have a very involved expression. Numerical analyses are much more enlightening. Figures \ref{cd106}, \ref{cd105} and \ref{cdp0} show some aspects from the numerical evaluation of Eq.~(\ref{stabcond}) for some specific scenarios.

One can see from Figs.~\ref{cd106} and \ref{cd105} that $C_2$ and $R$ play a relevant role into the stability of the system, unlike $C_1$. These results are expected since any change in $C_1$ would lead to physically similar configurations, with just a re-scaling of the outer region, while a modification of $C_2$ and $R$ would lead to an alteration in the normal flux of momentum through $\Sigma$, that clearly affects the stability. Besides, also the central pressure influences the stability of the system. This can be checked in Fig.~\ref{cdp0}. The aforementioned figures also show that fully general relativistic analyses may considerably change the classical picture, where the stability condition would merely read $[\rho]^+_-<0$. Indeed, as we can see in the specific case of $C_2=0.5$ (dashed curve) in Fig.~\ref{cd106}, there is the possibility of having some stable solutions with $[\rho]^+_->0$. The reason for this is the important role played by the pressure into the system, as well as by general relativistic corrections.

\subsection{Strange stars with crust}\label{strange}

We apply now the formalism to the especial case of \emph{strange quark stars}. In the core of astrophysical compact objects, such as neutron stars, the matter is expected to reach densities which are several times the nuclear saturation density, $\rho_{\rm nuc}\approx 3\times 10^{14}$~g~cm$^{-3}$. Calculations based on microscopic equations of state, which include only nucleonic degrees of freedom, show that the central densities of the most massive neutron stars can be of the order of $7$--$10\rho_{\rm nuc}$. In the traditional models, ordinary nuclear matter is assumed to be the true ground state of matter. However, it has been suggested that strange quark matter may be the authentic ground state of all matter \cite{1970Itoh,PhysRevD.4.1601,PhysRevD.30.272,Haensel:1986qb,1986ApJ...310..261A}. According to the strange matter hypothesis, the interior of neutron stars would be predominantly composed of up, down, and strange quarks, plus leptons that ensure the charge neutrality and the weak equilibrium of the star. These hypothetical compact stars composed of \emph{strange matter} are referred to as strange stars. Being self-bound by the strong force, the quark pressure in strange stars vanishes at a finite value of the energy-density, leading to the formation of a sharp surface. Since the electrons that guarantee the neutrality are blind to the strong force, they actually leak out from this sharp surface creating a thin electron-layer of a few hundred fermi in which a strong electric field exists. It has been suggested (see, e.g., \cite{1986ApJ...310..261A}) that strange stars not need to be necessarily bare cores as the ones just described but, instead, they can support, above the electronic layer, a crust of ordinary matter similar to outer crust of neutron stars. Therefore, owing to the very small size of the transition interface between the strange matter core and the crust in a strange star, we can analyze it within the thin-shell formalism and assess its stability.

In order to have some insights about this interesting case, we start our analyses with its simplest microscopic quark matter model, the MIT bag model~\cite{PhysRevD.9.3471}. Such a model assumes
that quarks constitutes a free Fermi gas inside a ``bag'' whose width is related to the value of the energy-density at which the pressure vanishes, i.e., the vacuum energy-density. In the limit of vanishing strange quark mass, $m_s\to 0$, the equation of state reduces to the simple linear expression \cite{1986ApJ...310..261A}:
\begin{equation}\label{bagmodel}
p=\frac{1}{3}(\rho - 4B),
\end{equation}
where $B$ is the bag constant. As in Ref.~\cite{1986ApJ...310..261A}, we shall adopt, without loss of generality, $B=(145~{\rm MeV})^4$. The precise value of the bag constant does not change the main qualitative conclusions to be drawn here.

We are interested here in investigating whether or not it is possible for the strange star to have a crust on top of its core's surface (with zero quark pressure). The density of the crust at the edge with the electronic layer has to be necessarily lower than the neutron-drip value, $\rho_{\rm drip}\approx 4.3\times 10^{11}$~g~cm$^{-3}$, since having zero electric charge, any free neutron created in the crust would flow to the core where it would be converted into strange matter. For the crust matter, we use the Baym-Pethick-Sutherland equation of state \cite{baym71b}.

\begin{figure}[!htbp]
	\centering
		\includegraphics[width=1.\hsize]{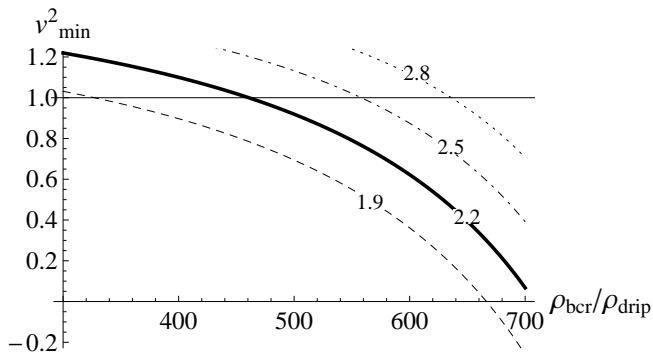}
		\caption{$v^2_{min}$ ``minimum square velocity'' as a function of the ``density at the base of the crust'', $\rho_{bcr}$, normalized by $\rho_{\rm drip}\approx 4.3\times 10^{11}$~g~cm$^{-3}$. Negative values for $v^2_{min}$ mean that any equation of state leads to stable solutions. Each curve is related to a given central density, in units of the nuclear density, $\rho_{\rm nuc}\approx 2.7\times 10^{14}$~g~cm$^{-3}\approx 628 \rho_{\rm drip}$.  From Eq.~(\ref{bagmodel}), one sees that the pressure vanishes at the density $4B \approx 1.51\rho_{\rm nuc}\approx 948\rho_{\rm drip}$. The crust was matched to the core exactly when the aforesaid density is reached for each case analyzed. The densities at the base of the crust that would render stable solutions to the thin-shells are only of the order of hundreds of the neutron drip density. Such densities are not admissible for crusts on quark cores.}\label{vmin}
\end{figure}

We solved the Einstein's equations (\ref{g11equ})--(\ref{derg00}) for the above described equation of state, for selected values of the central density and different densities at the base of the crust, which we denoted to as $\rho_{bcr}$, in each case. Then, we seek for values of the parameter $\eta$ that satisfy the stability condition (\ref{stabcond})
of the shell's effective potential. Our numerical results show that thin-shells splitting the quark phases from crusts are always unstable for 
densities at the base of the crust of the order of the neutron drip density. In Fig.~\ref{vmin} one can indeed verify that a quark star would be stable 
just if its density at the base of the crust was some hundred times the neutron drip one, which is clearly not permissible in the strange star hypothesis 
recalled above.

The previous result suggests us that strange stars should just have a tiny a crust cloaking the quark's core surface. Therefore, we should seek for solutions where the quark core would be matched directly with the Schwarzschild exterior spacetime, hence treating the crust itself as part of the thin-shell.  We recall that the quark stars we are analyzing here have at their core's edge null pressures and $\rho= 4B$. For this case and matching with Schwarzschild's solutions, we already know from section~\ref{constpress} that the associated thin-shells are stable, irrespective of the fluids they host. Physically speaking this result means that extremely thin crusts could always be taken as parts of thin-shells. Table~\ref{tab:widthquark} suggests that, indeed, very low densities at the base of the crusts would allow us to interpret the crusts as constituents of thin-shells.

\begin{table*}[!htbp]
\caption{\label{tab:widthquark}Width of the crusts for various strange stars configurations and densities at the base of the crust.}
\begin{ruledtabular}
\begin{tabular}{c c c}
$\rho_{bcr}/\rho_{\rm drip}$ & $\Delta R_{\rm crust}/R$ & $M_{\rm crust}/M_{\rm core}$ \\
\cline{1-3}
\multicolumn{3}{c}{$\rho_c=3.5\rho_{\rm nuc}$,\quad $R_c= 11.42$~km,\quad $M_c= 1.77 M_\odot$}\\
\cline{1-3}
\\
$10^{-9}$ & $2.88\times10^{-7}$ & $1.25\times10^{-16}$ \\
$10^{-5}$ & $3.49\times10^{-4}$ & $2.97\times10^{-12}$ \\
$10^{-2}$ & $6.06\times10^{-3}$ & $3.50\times10^{-8}$ \\
$10$ & $4.56\times10^{-3}$ & $4.65\times10^{-5}$ \\
\cline{1-3}
\multicolumn{3}{c}{$\rho_c=5.0\rho_{\rm nuc}$,\quad $R_c= 11.34$~km,\quad $M_c= 1.97 M_\odot$}\\
\cline{1-3}
\\
$10^{-9}$ & $2.32\times10^{-7}$ & $1.13\times10^{-16}$ \\
$10^{-5}$ & $2.81\times10^{-4}$ & $2.10\times10^{-12}$ \\
$10^{-2}$ & $4.87\times10^{-3}$ & $2.48\times10^{-8}$ \\
$10$ & $3.67\times10^{-3}$ & $3.28\times10^{-5}$ \\
\cline{1-3}
\multicolumn{3}{c}{$\rho_c=8.0\rho_{\rm nuc}$,\quad $R_c= 10.85$~km,\quad $M_c= 2.02 M_\odot$}\\
\cline{1-3}
\\
$10^{-9}$ & $2.01\times10^{-7}$ & $2.20\times10^{-16}$ \\
$10^{-5}$ & $2.43\times10^{-4}$ & $1.56\times10^{-12}$ \\
$10^{-2}$ & $4.21\times10^{-3}$ & $1.84\times10^{-8}$ \\
$10$ & $3.17\times10^{-3}$ & $2.44\times10^{-5}$
\end{tabular}
\end{ruledtabular}
\end{table*}

\section{Extension to slow rotation}\label{slowrot}

We turn now to show that our results remain unchanged even in the case where the shell is allowed to have a small rotation. This is indeed what one expects and we show it here just for self-consistency. For this case, one supposes that
\begin{equation}
ds_{\pm}^2=ds^2_{\pm}(s.s.)-2f_{\pm}(r_{\pm})r^2_{\pm}\sin^2\theta_{\pm} a_{\pm} dt_{\pm}d\varphi_{\pm}\label{smallrot},
\end{equation}
where $a_{\pm}$ are the rotation parameters in the regions $M_{\pm}$. Besides, the first term on the right hand side of Eq.~(\ref{smallrot}) is simply a shortcut for the spherically symmetric line element given by Eq.~(\ref{linelsph}). We take $\tilde\Sigma$, the hypersurface for this slow rotation case, in first order of approximation to be also spherically symmetric [$ r_{\pm} = R(\tau)$], but at this time we adapt on it the coordinates $\tilde y^{a}=(\tau,\tilde \theta, \tilde \varphi)$, such that
\begin{equation}
t_{\pm}=T_{\pm}(\tau),\;\; \varphi_{\pm}=\tilde \varphi + C_{\pm}(\tau),\;\; \theta_{\pm}= \tilde \theta \label{coordchange}.
\end{equation}
Then, it can be readily demonstrated that the geometry of $\tilde\Sigma$ is well defined (unique) if
\begin{equation}
\frac{d}{d\tau}C_{\pm}={f_{\pm}(R)\dot{T}_{\pm}}a_{\pm},\;\;
\dot{T}_{\pm}=\sqrt{e^{-2\beta_{\pm}}+\dot{R}^2}\, e^{\beta_{\pm}-\alpha_{\pm}}\label{coordcond}.
\end{equation}
The above conditions guarantee that the geometry of $\tilde\Sigma$ is
\begin{equation}
d\tilde s^2|_{\tilde\Sigma}=\tilde h_{ab}d\tilde y^a d\tilde y^b=-d\tau^2+R^2(\tau)(d{\tilde{\theta}}^2+\sin^2\tilde\theta d\tilde\varphi^2)\label{elemsmallrot}.
\end{equation}
Undemanding calculations lead to the additional extrinsic curvature
\begin{equation}
K^{\pm}_{\tau \tilde{\varphi}}=e^{-\beta_{\pm}-\alpha_{\pm}}R^2f'_{\pm}\sin^2\tilde{\theta}\, a_{\pm}\label{extcurv03}.
\end{equation}
The diagonalization of the surface energy-momentum tensor in this case is done by solving the eigenvalue equation $S^{a}{}_{b}\tilde u^{b}=-\tilde{\sigma}\tilde u^{a}$. The unknown quantities here are $\tilde{\sigma}$ and $\tilde u^{a}$, complemented with the normalization condition $\tilde u^{a}\tilde u_{a}=-1$. Besides, $\tilde u^{a}$ are the components of the tetrad decomposition of the four-velocity of the shell with respect to the coordinate system $\tilde y^a$. The pressure in the surface energy-momentum tensor in the $\tilde y^{a}$ coordinate system is simply $2\tilde{\cal P}=S^{ab}(\tilde u_a \tilde u_b+\tilde h_{ab})$. The solution to the above eigenvalue problem is
\begin{equation}
\tilde{\sigma}=\sigma,\;\;\tilde{\cal P}= {\cal P},\;\; \tilde u^{a}=(1,0,\omega),\;\; \omega=-\frac{S^{\tilde{\varphi}}{}_{\tau}}{\sigma+{\cal P}}\label{eigenveq}.
\end{equation}
Notice that $\omega$ is polar angle dependent. Eqs.~(\ref{eigenveq}) and (\ref{coordchange}) tell us that inertial observers inside the shell are rotating with
angular velocity proportional to $dC_+/d\tau$ with respect to the fixed stars when the inner spacetime is spherically symmetric. For the shell itself,
$\Omega_{shell}\propto\omega+ dC_+/d\tau$. At first order, there is no change in the parameters of the shell.  As we commented formerly, this is already accounted since the corrections imprinted by the rotation to the shell parameters must be independent of its direction of rotation. Nevertheless, up to first order of correction on the rotational parameters $a_{\pm}$, a frame dragging effect is present, whose associated angular velocity gives a direct information about the surface energy-momentum tensor parameters. We shall elaborate upon these issues in a forthcoming publication.

\section{Conclusions and discussion}\label{disc}


In the scope of the thin-shell formalism applied to stars, we have showed that whenever one considers phase transitions at constant pressures and with a negligible masses on the surfaces splitting them, the latter ones are always stable. This is relevant for commonly implemented phase transitions based on either the Maxwell or Gibbs constructions, since this would justify their use within the thin-shell formalism. In this case the degrees of freedom on the surface of discontinuity present in the dynamic case would always lead to stable configurations. Our analyses also show that only tiny crusts, associated with thin shells, could envelop on the surfaces of quark stars (at zero pressure). Nevertheless, whenever the match between the core and the crust is not done at the strange star's surface (where the quark pressure is null), it is always possible for the system to harbor thick crusts. This is so due to the steep increase of the quark pressure inwards, which would always allow a stable glue of the core with a crust at a radial position where their pressures equal, as we have showed previously.

When perturbations in the phases are also present, in principle they would also be dependent upon the surface degrees of freedom by means of additional boundary conditions to be taken for the stability problem, to be properly defined, what would also change the set of eigenfrequencies of the system. This will be investigated in a forthcoming publication. At this first approach, the aforementioned subtleties were not taken into account and we restricted ourselves to finding constraints where the surface perturbative analyses give a definite answer to the stability. This is due to the fact that the scenario where the phases are not perturbed evidences directly the consequences of the dynamics of the degrees of freedom of a shell, giving us thence insights for more elaborate analyses.

Concerning second and higher (even) order corrections to the rotational parameters of the shell in the stability analyses, a more detailed study is in order, to be attained elsewhere. Such a case could be relevant to assess the stability of millisecond pulsars. For the first order corrections to the rotational parameters, just frame-dragging effects are of relevance. If they were measured, then one could obtain a direct information of the shell parameters, that could shed some light into the issues raised in this work.

In addition to the simple example of quark matter analyzed in this work, there is the possibility of inducing conformal degrees of freedom into the transition hypersurface. This would be the case of the transition from hadronic (quark-confined) matter to color superconducting (color-flavor-locked, CFL) deconfined quark matter phase \cite{2001PhRvD..64g4017A,2008RvMP...80.1455A}, or in the case of the quantum Hall state between CFL and the hadronic phases \cite{2003PhLB..571...61I,2004PhLB..579..347I,2005PhRvD..71c4014I}. These systems lead to trace-free surface energy-momentum tensors, which in the spherically symmetric case imply ${\cal P} = \sigma/2$ [see Eq.~(\ref{Sab})]. Detailed stability analyses can be then done also in these cases once the phases associated with the transition hypersurface are given\footnote{It is worth stressing that the stability analyses done in Ref.~\cite{2007arXiv0710.3193C} are not correct since flux terms [Eq.~(\ref{Delta})], that are always present in continuous systems, were not taken into account there}. The location of the hypersurface clearly depends on the precise knowledge of the equation of state of the different phases. For instance, the transition CFL-hadronic hypersurface is located at a smaller radius with respect to the one considered here for the core-crust transition in a strange star, since the former transition occurs at higher matter densities.  Based on our results of section \ref{constpress}, we can conclude that, also in those more complex stratified stars, the stability of the hypersurfaces is guaranteed whenever the transition takes place at constant pressure. This is in contrast with the impossibility of having stable thin-shells in other context of linear thin-shell equations of state in the spherically symmetric case \cite{2002PhRvD..66h4021G}. This can be also derived from Eqs.~(\ref{evolequ}), (\ref{Delta}) [with $\Delta=0$], (\ref{shelldyn}) and (\ref{effpot}).

A very thought-provoking case that is possible to be analyzed in the Darmois-Israel formalism is the one where the surface energy-density satisfies $\sigma<0$. If this is valid, irrespective of its magnitude, then the inequality in Eq.~(\ref{stabcond}) should be reversed. Such a case would in principle render stable unstable configurations for the case $\sigma>0$. In this line, and in view of the very short distance scales involved in the interfaces, it is tempting to state that quantum-mechanical effects such as the Casimir one could be of some relevance there. Indeed, a simple calculation (using the expression for the energy-density to the Casimir effect for two concentric spheres, see e.g.~\cite{2008PhRvD..78f5023M}), shows that its energy-density is of the same order of magnitude as the Coulomb energy for a shell as the one present in the strange stars. This is an interesting issue that deserves to be better scrutinized.

Since there are good reasons for stars being stratified, surfaces degrees of freedom on surfaces of discontinuity could play a role there. It is then necessary to search for their observational fingerprints. In this regard, the ``glitches'' observed in pulsars could be a sign of the stratification of a system and deserve a closer look in light of the results presented in this work. The precession of the particle's orbits around a compact star could also give us information about surface/interface quantities, for instance related to the presence of a thin crust cloaking the core. The connection of this with the observed quasi-periodic oscillations (QPOs) for instance in low X-ray binaries \cite{1985Natur.316..225V} could be particularly relevant (see, e.g.,~\cite{pachon12}, and references therein).

\begin{acknowledgments}
J.A.R. is indebted to Professor Thibault Damour for insightful discussions on the subject of this work in the various occasions of the International Relativistic Astrophysics (IRAP) PhD-Erasmus Mundus Joint Doctorate Schools held in Nice. We are likewise grateful to Professor Luis Herrera Cometa for fruitful discussions. J.P.P. acknowledges the support given by the Erasmus Mundus Joint Doctorate Program within the IRAP PhD, under the Grant Number 2011--1640 from EACEA of the European Commission.  J.G.C. and J.A.R. acknowledge the support by the International Cooperation Program CAPES-ICRANet financed by CAPES -- Brazilian Federal Agency for Support and Evaluation of Graduate Education within the Ministry of Education of Brazil.
\end{acknowledgments}


\appendix
\section{Speed of the sound for shells embedded in continuous systems}
\label{app:sound}

In this section we use the $-2$ metric signature. In order to describe the stability of a dynamic thin-shell, one has to study the properties of perturbations (sound) propagating on $\Sigma$, as we have just shown by Eq.~(\ref{stabcond}). Since our shell is embedded in a continuous medium, the normal flux of momentum, the right-hand side of Eq.~(\ref{evoequ}), must be properly taken into account. For the spherically symmetric case this can be easily done, as we shall show in this appendix. The important points to be realized are that the geometry of $\Sigma$, given by Eq.~(\ref{Sgeom}), for each instant of time $\tau$, is flat and that the energy momentum of $\Sigma$ is those of a perfect fluid. Hence, we can work with Cartesian coordinates, come back to spherical ones at the end of
the calculation, and then suppress the radial coordinate for finding the dynamics of perturbations on $\Sigma$. From the above comments, we can pose the problem in the following form. Given
\begin{equation}
T^{\mu\nu}_{\hspace{.3cm},\,\nu}=f^{\mu}(x^{\nu}),\;\;T^{\mu\nu}=(\rho + p)u^{\mu}u^{\nu}-p\eta^{\mu\nu}\label{gensoundeq},
\end{equation}
with $f^{\mu}$ a given four-vector dependent upon the spacetime coordinates, we want to find the equation governing the evolution of the
perturbations on the pressure $p$, the energy-density $\rho$ and the velocity of the fluid $\vec{v}$, when the unperturbed solution for
the latter is zero.
From Eq.~(\ref{gensoundeq}), we have that Eq.~(\ref{gensoundeq}) can be split into
\begin{equation}
(\rho u^{\mu})_{,\mu}+p\,u^{\mu}_{\hspace{.1cm},\,\mu}=f_{\mu}u^{\mu}\label{scaeq}
\end{equation}
and
\begin{equation}
(\rho+p)u^{\nu}_{\;,\,\mu}u^{\mu} + (p_{,\mu}+f_{\mu})(u^{\mu}u^{\nu}-\eta^{\mu\nu})=0\label{eulereq}.
\end{equation}
The above equations admit a solution with $u^{\mu}=\delta^\mu_0$ iff
\begin{equation}
\rho_{,t}=f_0,\;\;p_{,i}=-f_{i}\label{statsol},
\end{equation}
where $i=1,2,3$. Now, let us suppose that $\tilde{u}^{\mu}=(1,\delta v^{i})$, ${\tilde{p}}=p+\delta p$, and ${\tilde{\rho}}=\rho+\delta \rho$,
where $p$ and $\rho$ are given by the solutions to Eqs.~(\ref{statsol}) and $\delta{\rho}$ and $\delta{p}$ are functions of the spacetime coordinates.
By putting $\tilde{\rho}$, $\tilde{p}$ and $\tilde{u}^{\mu}$ into Eqs.~(\ref{scaeq}) and (\ref{eulereq}), taking into account Eq.~(\ref{statsol}) and working up to first order of approximation in the $\delta$ terms,
one has
\begin{equation}
(\delta\rho)_{,t}+[(\rho+p)\delta v^{i}]_{,i}=0\label{pervel}
\end{equation}
and
\begin{equation}
[(\rho + p)\delta v^{i}]_{,t}+(\delta p)^{,i}=0\label{perpres},
\end{equation}
where we defined $(\delta p)^{,i}\doteq \partial (\delta p)/\partial x^{i}$. Eqs.~(\ref{pervel}) and (\ref{perpres}) lead us to
\begin{equation}
(\delta \rho)_{,tt}-[(v_c^2\delta \rho)^{,i}]_{,i}=0\label{soundseq},
\end{equation}
where are assumed that
\begin{equation}
\delta p = \frac{\partial p}{\partial \rho}\delta \rho \doteq v_c^2\delta\rho \label{adiabcond}.
\end{equation}
In other words, we considered the system to be adiabatic. By assuming that $v_c=v_c(x^{\mu})$, we have that Eq.~(\ref{soundseq}) reads
\begin{equation}
\frac{\partial^2\delta\rho}{\partial t^2}-(\nabla^2v_c^2)\delta\rho -v_c^2\nabla^2\delta\rho -\delta^{ij}\partial_iv_c^2\partial_j\partial\delta\rho=0\label{expsoundexp},
\end{equation}
with $\nabla^2\doteq \delta^{ij}\partial_i\partial_j$ and $\delta^{ij}=-\eta^{ij}$. We see from the above equation that in general there will be a
damping factor for the propagation of disturbances.

From Eq.~(\ref{stabcond}), we need to analyze the speed of the sound at the equilibrium point of the shell [critical point of the effective potential in Eq.~(\ref{effpot})] in order to assess its stability. In order to analyze these propagations on the surface of $\Sigma$, we should leave out the $r$ coordinate of Eq.~(\ref{expsoundexp}), taking it as a constant, keeping just the spherical ones. By the spherical symmetry of the system, it is clear that $v_c$ is neither dependent on $\theta$ nor $\varphi$. Hence, from Eq.~(\ref{expsoundexp}), we conclude that the usual expression for the speed of the sound is the one to be used in our stability analyses.

\end{document}